\documentclass[english,preprint,onecolumn,aps,pre,eqsecnum,longbibliography,showpacs,showkeys,a4paper]{revtex4-1}
\usepackage[latin9]{inputenc}
\usepackage{color}
\definecolor{note_fontcolor}{rgb}{0.80078125, 0.80078125, 0.80078125}
\usepackage{babel}
\usepackage{prettyref}
\usepackage{mathrsfs}
\usepackage{amsmath}
\usepackage{amssymb}
\usepackage{esint}
\usepackage[unicode=true,
 bookmarks=true,bookmarksnumbered=true,bookmarksopen=true,bookmarksopenlevel=1,
 breaklinks=true,pdfborder={0 0 0},backref=false,colorlinks=true]
 {hyperref}
\hypersetup{
 pdfauthor={AINS}}

\makeatletter

\@ifundefined{textcolor}{}
{%
 \definecolor{BLACK}{gray}{0}
 \definecolor{WHITE}{gray}{1}
 \definecolor{RED}{rgb}{1,0,0}
 \definecolor{GREEN}{rgb}{0,1,0}
 \definecolor{BLUE}{rgb}{0,0,1}
 \definecolor{CYAN}{cmyk}{1,0,0,0}
 \definecolor{MAGENTA}{cmyk}{0,1,0,0}
 \definecolor{YELLOW}{cmyk}{0,0,1,0}
 }
\numberwithin{equation}{section}
\numberwithin{figure}{section}
\numberwithin{table}{section}

\usepackage{graphicx}

\DeclareGraphicsExtensions{.png,.pdf,.eps}
\graphicspath{{pictures/}}

\makeatother

\begin{document}

\title{Quantum features derived from the classical model of a bouncer-walker
coupled to a zero-point field}

\author{Herbert \surname{Schwabl}\textsuperscript{1}}

\email[E-mail: ]{ains@chello.at}

\homepage[Visit: ]{http://www.nonlinearstudies.at/}

\selectlanguage{english}%

\author{Johannes \surname{Mesa Pascasio}\textsuperscript{1,2}}

\email[E-mail: ]{ains@chello.at}

\homepage[Visit: ]{http://www.nonlinearstudies.at/}

\selectlanguage{english}%

\author{Siegfried \surname{Fussy}\textsuperscript{1}}

\email[E-mail: ]{ains@chello.at}

\homepage[Visit: ]{http://www.nonlinearstudies.at/}

\selectlanguage{english}%

\author{Gerhard \surname{Gr\"ossing}\textsuperscript{1}}

\email[E-mail: ]{ains@chello.at}

\homepage[Visit: ]{http://www.nonlinearstudies.at/}

\selectlanguage{english}%

\affiliation{\textsuperscript{1}Austrian Institute for Nonlinear Studies, Akademiehof\\
 Friedrichstr.~10, 1010 Vienna, Austria}

\affiliation{\textsuperscript{2}Institute for Atomic and Subatomic Physics, Vienna
University of Technology\\
Operng.~9, 1040 Vienna, Austria \vspace*{2cm}
}

\date{\today}
\begin{abstract}
In our bouncer-walker model a quantum is a nonequilibrium steady-state
maintained by a permanent throughput of energy. Specifically, we consider
a ``particle'' as a bouncer whose oscillations are phase-locked
with those of the energy-momentum reservoir of the zero-point field
(ZPF), and we combine this with the random-walk model of the walker,
again driven by the ZPF. Starting with this classical toy model of
the bouncer-walker we were able to derive fundamental elements of
quantum theory~\cite{Groessing.2011explan}. Here this toy model
is revisited with special emphasis on the mechanism of emergence.
Especially the derivation of the total energy $\hbar\omega_{0}$ and
the coupling to the ZPF are clarified. For this we make use of a sub-quantum
equipartition theorem. It can further be shown that the couplings
of both bouncer and walker to the ZPF are identical. Then we follow
this path in accordance with Ref.~\cite{Groessing.2010emergence},
expanding the view from the particle in its rest frame to a particle
in motion. The basic features of ballistic diffusion are derived,
especially the diffusion constant $D$, thus providing a missing link
between the different approaches of our previous works \cite{Groessing.2011explan,Groessing.2010emergence}.%

\global\long\def\VEC#1{\mathbf{#1}}
\global\long\def\d{\,\mathrm{d}}
\global\long\def\e{{\rm e}}
\global\long\def\meant#1{\left<#1\right>}
\global\long\def\meanx#1{\overline{#1}}
\global\long\def\mpbracket{\ensuremath{\genfrac{}{}{0pt}{1}{-}{\scriptstyle (\kern-1pt +\kern-1pt )}}}
\global\long\def\pmbracket{\ensuremath{\genfrac{}{}{0pt}{1}{+}{\scriptstyle (\kern-1pt -\kern-1pt )}}}
\global\long\def\p{\partial}

\end{abstract}


\maketitle

\section{Introduction}

\label{sec:intro}

As explicated already in some of our previous papers \cite{Groessing.2011explan,Groessing.2010emergence,Groessing.2011dice,Groessing.2012doubleslit},
we understand the quantum as a well-coordinated \textit{emergent system},
where particle-like and wave-like phenomena are the result of both
stochastic and regular dynamical processes which exchange energy with
the surrounding ``vacuum'', i.e., the zero-point field (ZPF). Thus,
a quantum is modeled as a nonequilibrium steady-state maintained by
a permanent throughput of energy. Specifically, we consider a ``particle''
as a bouncer-walker whose combined movements are coupled to the energy-momentum
reservoir of the ZPF. The notion of the bouncer-walker is derived
from classical physics, as shown experimentally via the ``bouncing
droplets'' of Couder's group \cite{Couder.2005,Couder.2006single-particle,Protiere.2006,Eddi.2009,Fort.2010path-memory}.

The possible existence of such a corresponding, underlying ``medium''
is \textit{a priori} independent of quantum theory. For similar views,
compare, for example, the works of Hestenes~\cite{Hestenes.2009zitterbewegung},
Khrennikov~\cite{Khrennikov.2011prequantum}, Nelson~\cite{Nelson.1985quantum},
Nottale~\cite{Nottale.2011scale}, and de~la~Pe\~na and Cetto~\cite{Pena.2011}.
We want to underline that by using the expression ``classical'',
we imply the ``updated'' present-day status of classical physics,
i.e., including present-day statistical physics, nonequilibrium thermodynamics,
and the like. Vacuum fluctuations in our terminology thus refer to
the statistical mechanics of the ZPF, i.e., a ``classical'' sub-quantum
medium.

In our picture the quantum emerges via phase coupling of the bouncer's
oscillatory frequency $\omega$ with corresponding modes of the ZPF
fluctuations, combined with the random-walk model of the walker, again
driven by the ZPF. Starting with this toy model of the bouncer-walker,
we were able to derive fundamental elements of quantum theory from
such a classical approach. In \prettyref{sec:modeling} this toy model
is revisited according to~\cite{Groessing.2011explan}, but this
time with special emphasis on the mechanism of emergence. Especially,
the derivation of the total energy $\hbar\omega_{0}$ and the coupling
to the ZPF are clarified. \prettyref{sec:Gaussian} follows this path,
expanding the view from the particle in its rest frame to a particle
in motion~\cite{Groessing.2010emergence}. The basic features of
ballistic diffusion are derived via a link to the individual bouncer-walker
model.

\section{Modeling the quantum: ``bouncer'' and ``walker''\label{sec:modeling}}

The bouncer is modeled as a classical oscillator with the following
Newtonian equation 
\begin{equation}
m\ddot{\mathsf{x}}=-m\omega_{0}^{2}\mathsf{x}-2\gamma m\mathsf{\dot{x}}+F_{0}\cos\omega t.\label{eq:2.1}
\end{equation}
 Eq.~(\ref{eq:2.1}) describes a forced oscillation of a mass $m$
swinging around a center point along $\mathscr{\mathsf{x}}(t)$ with
amplitude $A$ and resonant frequency $\omega_{0}$. The damping factor
$\gamma$, or friction, allows the oscillator to exchange energy to
the ZPF, which in this case is modeled by a locally independent driving
force $F(t)=F_{0}\cos\omega t$. In the center of mass frame, the
system is characterized by a single degree of freedom (DOF).

The stationary solution of Eq.~(\ref{eq:2.1}), i.e., for $t\gg\gamma^{-1}$,
with the ansatz 
\begin{equation}
\mathsf{x}(t)=A\cos(\omega t+\varphi),\label{eq:2.2}
\end{equation}
 yields for the phase shift between the forced oscillation and the
forcing oscillation that 
\begin{equation}
\tan\varphi=-\frac{2\gamma\omega}{\omega_{0}^{2}-\omega^{2}}\;,\label{eq:2.3}
\end{equation}
 and for the amplitude of the forced oscillation 
\begin{equation}
A(\omega)=\frac{F_{0}/m}{\sqrt{(\omega_{0}^{2}-\omega^{2})^{2}+(2\gamma\omega)^{2}}}\;.\label{eq:2.4}
\end{equation}

It is well known that the system is stable at the resonant frequency
\textit{$\omega_{0}$ }\textit{\emph{of the free undamped oscillator}}
\begin{equation}
\omega=\omega_{0}.\label{eq:2.11-1}
\end{equation}
 With identity (\ref{eq:2.11-1}) we introduce the notations 
\begin{equation}
\tau=\frac{2\pi}{\omega_{0}}\;,\quad r:=A(\omega_{0})=\frac{F_{0}}{2\gamma m\omega_{0}}\;.\label{eq:2.12-1}
\end{equation}

We assume the net energy balance of the exchange oscillator--ZPF to
be zero, i.e., the oscillator--ZPF system is in a steady state. For
this, we analyze the energetic balance. By multiplying Eq.~(\ref{eq:2.1})
with $\dot{\mathsf{x}}$ we obtain
\begin{align}
m\mathsf{\ddot{x}}\mathsf{\dot{x}}+m\omega_{0}^{2}\mathsf{x\dot{x}}=-2\gamma m\dot{\mathsf{x}}^{2}+F_{0}\cos(\omega t)\dot{\mathsf{x}}.\label{eq:2.5}
\end{align}
Due to the friction term $-2\gamma m\dot{\mathsf{x}}^{2}$ the oscillator
loses its energy to the ZPF bath, whereas the oscillator regains its
power $F_{0}\cos(\omega t)\dot{\mathsf{x}}$ from the energy bath
via the force $F(t)$. We can rewrite Eq.~(\ref{eq:2.5}) as
\begin{align}
\frac{{\rm d}}{{\rm d}t}\left(\frac{1}{2}m\dot{\mathsf{x}}^{2}+\frac{1}{2}m\omega_{0}^{2}\mathsf{x}^{2}\right)=-2\gamma m\dot{\mathsf{x}}^{2}+F_{0}\cos(\omega t)\dot{\mathsf{x}}=0.\label{eq:2.6}
\end{align}
 The term within the brackets is the Hamiltonian of the system. By
inserting Eqs.\eqref{eq:2.2} and \eqref{eq:2.11-1} we see 
\begin{equation}
\mathcal{H}=\frac{1}{2}m\dot{\mathsf{x}}^{2}+\frac{1}{2}m\omega_{0}^{2}\mathsf{x}^{2}=\frac{m\omega_{0}^{2}A^{2}}{2}=\text{const.},\label{eq:2.14}
\end{equation}
 thus providing the vanishing of Eq.~\eqref{eq:2.6}.

One can write down the net work-energy that is taken up by the bouncer
during each period $\tau$ as 
\begin{align}
W_{{\rm bouncer}} & =\intop_{\tau}F_{0}\cos(\omega t)\dot{\mathsf{x}}\d t=\intop_{\tau}2\gamma m\dot{\mathsf{x}}^{2}\d t\nonumber \\
 & =2\gamma m\omega^{2}A^{2}\intop_{\tau}\sin^{2}(\omega t+\varphi)\d t\nonumber \\
 & =\gamma m\omega^{2}A^{2}\tau.\label{eq:2.7}
\end{align}
With Eqs.~(\ref{eq:2.11-1}) and (\ref{eq:2.12-1}) we obtain in
the steady state 
\begin{equation}
W_{{\rm bouncer}}=W_{{\rm bouncer}}(\omega_{0})=\gamma m\omega_{0}^{2}r^{2}\tau=2\pi\gamma m\omega_{0}r^{2}.\label{eq:2.13}
\end{equation}
We now shift to polar coordinates, which allow us to model the system
in continuous circular motion. If one introduces the angle $\theta(t):=\omega_{0}t$
and substitutes Eq.~(\ref{eq:2.2}) into Eq.~(\ref{eq:2.14}), this
yields, as is well known, the two equations 
\begin{align}
\ddot{r}-r\dot{\theta}^{2}+\omega_{0}^{2}r & =0,\label{eq:2.15}
\end{align}
 and 
\begin{align}
r\ddot{\theta}+2\dot{r}\dot{\theta} & =0.\label{eq:2.16}
\end{align}
 From Eq.~(\ref{eq:2.16}), an invariant quantity is obtained: it
is the angular momentum 
\begin{equation}
L(t)=mr^{2}\dot{\theta}(t).\label{eq:2.17}
\end{equation}
 With $\theta(t)=\omega_{0}t$, and thus $\dot{\theta}=\omega_{0}$,
the quantity of Eq.~(\ref{eq:2.17}) becomes a time-invariant expression,
which we denote as 
\begin{equation}
\hbar:=mr^{2}\omega_{0}.\label{eq:2.18}
\end{equation}
 Note that $L(t)$ is an invariant even more generally, i.e., for
$\theta(t):=\int\omega(t)\d t$. Still, for all cases where the time
average $\meant{\theta(t)}=\omega_{0}t$, one can again write down
$\hbar$ in the form Eq.~\eqref{eq:2.18}. Thus, we rewrite our result
(\ref{eq:2.13}) as 
\begin{equation}
W_{{\rm bouncer}}=2\pi\gamma\hbar.\label{eq:2.19}
\end{equation}

Now let us revisit the energy \eqref{eq:2.14} of the linear harmonic
oscillator, which reads, together with Eqs.~\eqref{eq:2.12-1} and
\eqref{eq:2.18},
\begin{equation}
\mathcal{H}=\frac{m\omega_{0}^{2}r^{2}}{2}=\frac{\hbar\omega_{0}}{2}=E_{\mathrm{bouncer}}.\label{eq:2.20}
\end{equation}

In a second step, we model another dissipative system. The ``walker''
is a ``particle'' driven via a stochastic force, e.g., due to not
just one regular, but due to different fluctuating wave-like configurations
in the ZPF environment. The particle's motion, which will generally
assume a Brownian-type character, is then described (in any one dimension)
by a Langevin stochastic differential equation with velocity $u=\dot{x}$,
driving force $f(t)$, and friction coefficient $\zeta$, 
\begin{equation}
m\dot{u}=-m\zeta u+f(t).\label{eq:3.1}
\end{equation}
Again, ``friction'', earlier represented by $\gamma$ and now by
$\zeta$, generally describes the coupling between the oscillator
(or particle in motion) on the one hand, and the ZPF bath on the other
hand. The time-dependent force $f(t)$ is stochastic, i.e., one has
as usual for the time-averages 
\begin{equation}
\meant{f(t)}=0\;,\quad\meant{f(t)f(t')}=\phi(t-t'),\label{eq:3.2}
\end{equation}
 where $\phi(t)$ differs noticeably from zero only for $t<\zeta^{-1}$.
The correlation time $\zeta^{-1}$ denotes the time during which the
fluctuations of the stochastic force remain correlated.

One usually introduces a coefficient $\lambda$ that measures the
strength of the mean square deviation of the stochastic force, such
that 
\begin{equation}
\phi(t)=\lambda\delta(t).\label{eq:3.7}
\end{equation}
 Since friction increases in proportion to the frequency of the stochastic
collisions, there must be a connection between $\lambda$ and $\zeta$.
One solves the Langevin equation (\ref{eq:3.1}) in order to find
this connection. Solutions of this equation are well known from the
Ornstein-Uhlenbeck theory of Brownian motion \cite{Uhlenbeck.1930theory,Coffey.2004}.

Since the dependence of $f(t)$ is known only statistically, one does
not consider the average value of $u(t)$, but instead that of its
square, 
\begin{equation}
\begin{aligned}\meanx{u^{2}(t)} & =\e^{-2\zeta t}\intop_{0}^{t}\d\tau\,\intop_{0}^{t}\d\tau'\e^{\zeta(\tau+\tau')}\phi(\tau-\tau')\frac{1}{m}+u_{0}^{2}\e^{-2\zeta t}\\
 & =\frac{\lambda}{2\zeta m^{2}}\left(1-\e^{-2\zeta t}\right)+u_{0}^{2}\e^{-2\zeta t}\quad\stackrel{t\gg\zeta^{-1}}{\longrightarrow}\quad\frac{\lambda}{2\zeta m^{2}}\;,
\end{aligned}
\label{eq:3.8}
\end{equation}
 with $u_{0}$ being the initial value of the velocity. For $t\gg\zeta^{-1}$,
the term with $u_{0}$ becomes negligible, i.e., $\zeta^{-1}$ then
plays the role of a relaxation time. We require that our particle
attains thermal equilibrium \cite{Groessing.2008vacuum,Groessing.2009origin}
after long times, so that due to the \textit{equipartition theorem
on the sub-quantum level} the average value of the kinetic energy
becomes 
\begin{equation}
\frac{1}{2}m\,\meanx{u^{2}(t)}=\frac{\lambda}{4\zeta m}=E_{{\rm zp}}=\frac{1}{2}kT_{0}.\label{eq:3.9}
\end{equation}
 where we introduce the average kinetic Energy $E_{{\rm zp}}$ of
the zero-point field, which is the sub-quantum analogon to the thermodynamical
expression $k_{{\rm B}}T/2$, where $k_{{\rm B}}$ is Boltzmann's
constant, and $T$ the classical temperature, whereas $T_{0}$ in
our scenario denotes the vacuum temperature. However, as we today
neither know $T_{0}$ nor the constant $k$ (unless it should turn
out as identical to $k_{{\rm B}}$), we shall mostly stick to formally
using $E_{{\rm zp}}$. In other words, we shall use the specification
``$kT_{0}$'' only occasionally, i.e., in order to point out the
close analogy to the usual thermodynamical formalism, and as a reminder
that $E_{{\rm zp}}$ is the ``kinetic temperature'' of the vacuum's
heat reservoir.

From Eq.~(\ref{eq:3.9}) one obtains an Einstein-type relation 
\begin{equation}
\lambda=4\zeta mE_{{\rm zp}}.\label{eq:3.10}
\end{equation}
 Similarly, we obtain the mean square displacement of $x(t)$ for
$t\gg\zeta^{-1}$. Therefore, one integrates twice to obtain 
\begin{equation}
\meanx{x^{2}(t)}=\intop_{0}^{t}\d\tau\intop_{0}^{t}\d\tau'\frac{\lambda}{2\zeta m^{2}}\e^{-\zeta|\tau-\tau'|}\simeq\frac{\lambda}{\zeta^{2}m^{2}}t=2Dt,\label{eq:3.11}
\end{equation}
 with the diffusion constant turning out as 
\begin{equation}
D=\frac{\lambda}{2\zeta^{2}m^{2}}=\frac{2E_{{\rm zp}}}{\zeta m}\;.\label{eq:3.12}
\end{equation}

Now we remind ourselves that we deal here with a steady-state system.
Just as with the friction $\zeta$ there exists a flow of (kinetic)
energy into the ZPF environment, there must also exist a work-energy
flow back into our system of interest. For its calculation, we multiply
Eq.~(\ref{eq:3.1}) by $u=\dot{x}$ and obtain an energy-balance
equation. With a natural number $n>0$ chosen so that $n\tau$ is
large enough to make all fluctuating contributions negligible, it
yields for the duration of time $n\tau$ the net work-energy of the
walker 
\begin{equation}
W_{{\rm walker}}=\intop_{n\tau}m\zeta\,\meanx{\dot{x}^{2}}\d t=m\zeta\intop_{n\tau}\meanx{u^{2}(t)}\d t.\label{eq:3.13}
\end{equation}
 Inserting Eq.~(\ref{eq:3.9}), we obtain 
\begin{equation}
W_{{\rm walker}}=n\tau m\zeta\,\meanx{u^{2}(t)}=2n\tau\zeta E_{{\rm zp}}.\label{eq:3.14}
\end{equation}
We have so far analyzed two perspectives for our model of a single-particle
quantum system: 
\begin{enumerate}
\item A harmonic oscillator is driven by the environment via a periodic
force $F_{0}\cos\omega_{0}t$. However, in the $N$-dimensional reference
frame of the laboratory, the oscillation is not fixed \textit{a priori}.
Rather, with $\hbar$ as angular momentum, there will be a free rotation
in all $N$ dimensions, and possible exchanges of energy will be equally
distributed in a stochastic manner. 
\item Concerning the latter, the flow of energy is on average distributed
evenly via the friction $\gamma$ in all $N$ dimensions of the laboratory
frame. It can thus also be considered as the stochastic source of
the particle moving in $N$ dimensions, each described by the Langevin
equation \eqref{eq:3.1}. 
\end{enumerate}
In order to make the result comparable with Eq.~(\ref{eq:2.19}),
we choose $\tau=2\pi/\omega_{0}$ to be identical with the period
of Eq.~\eqref{eq:2.12-1}. The work-energy for the particle undergoing
Brownian motion can thus be written as 
\begin{equation}
W_{{\rm walker}}=n\frac{4\pi}{\omega_{0}}\zeta E_{{\rm zp}},\label{eq:3.15}
\end{equation}
and, for the general case of $N$ DOF 
\begin{equation}
W_{{\rm walker}}=n\frac{N4\pi}{\omega_{0}}\zeta E_{{\rm zp}}.\label{eq:3.22}
\end{equation}
Accordingly, the walker gains its energy from the heat bath via the
oscillations of the bouncer-ZPF bath system in $N$ dimensions: The
bouncer, via the coupling $\gamma$, pumps energy to and from the
heat bath. There is a continuous flow from the bath to the oscillator,
and \textit{vice versa}. Moreover, and most importantly, during that
flow, for long enough times $n\tau$, this coupling of the bouncer
can be assumed to be exactly identical with the coupling of the walker.
For this reason, we directly compare the results of Eqs.~(\ref{eq:2.19})
and (\ref{eq:3.22}), 
\begin{equation}
nW_{{\rm bouncer}}=W_{{\rm walker}}.\label{eq:4.1}
\end{equation}
With $n\gg1$, since we have to take the mean over a large number
of stochastic motions, we get 
\begin{equation}
n2\pi\gamma\hbar=n\frac{N4\pi}{\omega_{0}}\zeta E_{{\rm zp}}.\label{eq:4.2}
\end{equation}

Now, one generally has that the total energy of a sinusoidal oscillator
exactly equals twice its average kinetic energy. With Eq.~\eqref{eq:2.20}
the bouncer model provided already
\begin{equation}
E_{{\rm tot}}=2E_{{\rm bouncer}}=\hbar\omega_{0},\label{eq:4.2-2}
\end{equation}
and we compare this result with Eq.~\eqref{eq:4.2}. We achieve the
same result for the total energy only if both systems, the bouncer
and the walker, are coupled with the same strength to the ZPF bath,
i.e., the friction coefficient for both the bouncer and the walker
must be identical, $\gamma=\zeta$. We have thus \textit{derived}
the total energy of our model for a quantum ``particle'', i.e.,
a driven steady-state oscillator system, as
\begin{equation}
E_{{\rm tot}}=2NE_{{\rm zp}}=2E_{{\rm bouncer}}=\hbar\omega_{0},\label{eq:4.3}
\end{equation}
where $\hbar$, as defined in Eq.~\eqref{eq:2.18}, can now be identified
with Planck's reduced constant. Note that via \eqref{eq:2.18}, $\hbar$
is defined as angular momentum in exactly the same manner as it is
obtained in an independent earlier derivation by Puthoff~\cite{Puthoff.1987ground}.

Now, with Boltzmann's relation $\Delta Q=2\omega_{0}\delta S$ between
the heat applied to an oscillating system and a change in the action
function $\delta S=\delta\int E_{{\rm kin}}\d t$, respectively, \cite{Groessing.2008vacuum,Groessing.2009origin}
one has 
\begin{equation}
\nabla Q=2\omega_{0}\nabla(\delta S).\label{eq:4.6}
\end{equation}
 $\delta S$ relates to the momentum fluctuation via 
\begin{equation}
\nabla(\delta S)=\delta\mathbf{p}=:m\mathbf{u}=-\frac{\hbar}{2}\frac{\nabla P}{P}\;,\label{eq:35}
\end{equation}
 and therefore, with $P=P_{0}\e^{-\delta Q/kT_{0}}$ and \eqref{eq:4.3},
\begin{equation}
m\mathbf{u}=\frac{\nabla Q}{2\omega_{0}}\;.\label{eq:4.8a}
\end{equation}
 As the friction force in Eq.~(\ref{eq:3.1}) is equal to the gradient
of the heat flux, 
\begin{equation}
m\zeta\mathbf{u}=\nabla Q,\label{eq:4.9}
\end{equation}
 comparison of \eqref{eq:4.8a} and \eqref{eq:4.9} provides now a
detailed expression for the coupling to the ZPF bath 
\begin{equation}
\zeta=\gamma=2\omega_{0}.\label{eq:4.10}
\end{equation}
 Note that with Eqs.~\eqref{eq:4.3} and \eqref{eq:4.10} one obtains
in one dimension the expression for the diffusion constant \eqref{eq:3.12}
as 
\begin{equation}
D=\frac{2E_{{\rm zp}}}{\zeta m}=\frac{\hbar}{2m}\;,\label{eq:4.12}
\end{equation}
 which is exactly the usual expression for $D$ in the context of
quantum mechanics.

Finally, we can also introduce the recently proposed concept of an
``entropic force'' \cite{Verlinde.2011origin,Padmanabhan.2010thermodynamical}.
We revisit Eq.~\eqref{eq:4.3} and look at the total energy equaling
a total work applied to the system. With $S_{{\rm e}}$ denoting the
entropy one can write 
\begin{align}
E_{{\rm tot}} & =2\,\meant{E_{{\rm kin}}}=:\VEC F\cdot\Delta\VEC x=T_{0}\Delta S_{{\rm e}}=\frac{1}{2\pi}\oint\nabla Q\cdot\d\VEC r\nonumber \\
 & =\Delta Q\,\text{(circle)}=2\left[\frac{\hbar\omega_{0}}{4}-\left(-\frac{\hbar\omega_{0}}{4}\right)\right]=\hbar\omega_{0}\label{eq:4.13}
\end{align}
which provides an ``entropic'' view of a harmonic oscillator in
its thermal bath.

We know already the total energy of a simple harmonic oscillator $E_{\mathrm{bouncer}}$,
Eq.~\eqref{eq:2.20}. The average kinetic energy of a harmonic oscillator
is given by half of its total energy, i.e., by $\meant{E_{{\rm kin}}}=E_{\mathrm{bouncer}}/2=\hbar\omega_{0}/4$,
which --- because of the local equilibrium --- is both the average
kinetic energy of the bath and that of the ``bouncer'' particle.
As the latter during one oscillation varies between $0$ and $\hbar\omega_{0}/2$,
one has the following entropic scenario. When it is minimal, the tendency
towards maximal entropy will provide an entropic force equivalent
to the absorption of the heat quantity $\Delta Q=\hbar\omega_{0}/4$.
Similarly, when it is maximal, the same tendency will now enforce
that the heat $\Delta Q=\hbar\omega_{0}/4$ is given off again to
the ``thermostat'' of the thermal bath. In sum, then, the total
energy throughput $E_{{\rm tot}}$ along a full circle will equal,
according to Eq.~(\ref{eq:4.13}), $2\meant{E_{{\rm kin}}}({\rm circle})=2\hbar\omega_{0}/2=\hbar\omega_{0}$.
In other words, the formula $E=\hbar\omega_{0}$ does not refer to
a classical ``object'' oscillating with frequency $\omega_{0}$,
but rather to a process of a ``fleeting constancy'': due to entropic
requirements, the energy exchange between bouncer and heat bath will
constantly consist of absorbing and emitting heat quantities such
that in sum the ``total particle energy'' emerges as $\hbar\omega_{0}$.

As was shown in \cite{Groessing.2011explan} one can continue along
these classical lines to express further features of quantum mechanics,
like the energy spectrum of the harmonic oscillator, or spin, for
example.

\section{Gaussian particle behaviour\label{sec:Gaussian}}

So far, we have described the emerging entity in its rest frame. Now
we expand this model and describe the randomly moving walker, characterized
by the diffusion constant $D$, Eq.~(\ref{eq:4.12}). Here we follow
the arguments presented in~\cite{Groessing.2010emergence}. As shown
above, the nonequilibrium steady-state is characterized by a permanent
throughput of energy, or heat flow. First, let us reconsider the Brownian
motion of the particle, but from another perspective as compared to
the previous chapter. The Brownian motion is a form of kinetic energy
provided by the ZPF interaction, and is of course different from the
\textquotedblleft{}ordinary\textquotedblright{} kinetic energy of
the particle, which will be introduced later. The total energy of
the whole system can be written as
\begin{equation}
E_{{\rm tot}}=\hbar\omega+\frac{\left(\delta p\right)^{2}}{2m}\:,\label{eq:3.1-1}
\end{equation}
where $\hbar\omega$ is the generalized total energy of the particle
and $\delta p:=mu$ is said additional, fluctuating momentum of the
particle of mass $m$ \cite{Groessing.2010emergence}. Note that $\delta p$
can take on an arbitrary value such that $E_{\mathrm{tot}}$ is generally
variable.

Every bouncer-walker is a rapidly oscillating object, which itself
is guided by the ZPF environment that also contributes some fluctuating
momentum to the walker\textquoteright{}s propagation. In fact, the
walker is the cause of the waves surrounding the particle, and the
detailed structure of the wave configurations influences the walker\textquoteright{}s
path, as the particle both absorbs heat from and emits heat into its
environment, both cases of which can be described in terms of momentum
fluctuations. Thus, if we imagine the bouncing of a walker in its
\textquotedblleft{}fluid\textquotedblright{} environment, the latter
will become \textquotedblleft{}excited\textquotedblright{} or \textquotedblleft{}heated
up\textquotedblright{} wherever the momentum fluctuations direct the
particle to. After some time span (which can be rather short, considering
the very rapid oscillations of elementary particles), a whole area
of the particle\textquoteright{}s environment will be coherently heated
up in this way.

Now we expand this further to a source of identical particles, which
are prepared in such a way that each one ideally has an initial (classical)
velocity $\mathbf{v}$, which is also called ``convective'' velocity.
Similar arguments are presented by Gr\"ossing~\cite{Grossing.2012quantum}
and~\cite{Groessing.2011explan,Groessing.2010emergence,Groessing.2011dice,Groessing.2012doubleslit,Groessing.2008vacuum,Groessing.2009origin,Groessing.2010entropy},
but here we want to focus on the interaction with the ZPF. The particles
emerge from the source, one at a time only, with a Gaussian probability
density $P$. This comes along with a heat distribution generated
by the oscillating (\textquotedblleft{}bouncing\textquotedblright{})
particle(s), with a maximum at the center of the aperture $\boldsymbol{x}_{0}=\boldsymbol{\mathbf{v}}t$.
In one dimension the corresponding solution of the heat equation is
then
\begin{equation}
P\left(x,t\right)=\frac{1}{\sqrt{2\pi}\sigma}e^{-\frac{\left(x-x_{0}\right)^{2}}{2\sigma^{2}}}\:,\label{eq:42}
\end{equation}
with the usual variance $\sigma^{2}=\meanx{\left(\Delta x\right)^{2}}=\meanx{\left(x-x_{0}\right)^{2}}$,
where we shall choose $x_{0}\left(t=0\right)=0$. Note that from Eq.~(\ref{eq:3.1-1})
one has for the averages over particle positions and fluctuations
\begin{equation}
\meanx{E_{\mathrm{tot}}}=\meanx{\hbar\omega}+\frac{\meanx{\left(\delta p\right)^{2}}}{2m}=\text{const.},\label{eq:3.3}
\end{equation}
with the mean values (generally defined in $N$-dimensional configuration
space
\begin{equation}
\meanx{\left(\delta p\right)^{2}}:=\int P\left(\delta p\right)^{2}\d^{N}x\:.\label{eq:3.4}
\end{equation}
Eq.~(\ref{eq:3.3}) is a statement of total average energy conservation,
i.e., holding for all times $t$ . Being a central argument, it is
also the starting point for~\cite{Groessing.2010emergence}. In Eq.~(\ref{eq:3.3}),
a variation in $\delta p$ implies a varying \textquotedblleft{}particle
energy\textquotedblright{} $\hbar\omega$, and \textit{vice versa},
such that each of the summands on the right hand side for itself is
not conserved. As can be shown~\cite{Groessing.2010emergence,Grossing.2012quantum},
there is an exchange of momentum between the two terms providing a
net balance
\begin{equation}
m\delta v-m\delta u=0\label{eq:3.5.}
\end{equation}
where $\delta v$ describes a change in the convective velocity $v$
paralleled by the \textquotedblleft{}diffusive\textquotedblright{}
momentum fluctuation $\delta(\delta p):=m\delta u$ in the thermal
environment.

Now let us look at the contributions of the diffusive and convective
velocities to the total energy. As from Eq.~(\ref{eq:3.3}) one has
that $\frac{\partial}{\partial t}\meanx{E_{\mathrm{tot}}}=0$ and
thus also $\delta\meanx{E_{\mathrm{tot}}(t)}-\delta\meanx{E_{\mathrm{tot}}(0)}=0$,
and as only the kinetic energy varies, one obtains $\delta\meanx{E_{\mathrm{kin}}(t)}=\delta\meanx{E_{\mathrm{kin}}(0)}=\mathrm{const}.$,
which yields for any $t$, with \eqref{eq:35} and \eqref{eq:42},
\begin{equation}
\delta\meanx{E_{\mathrm{kin}}(t)}=\frac{m}{2}\meanx{\left(\delta v\right)^{2}}+\frac{m}{2}\meanx{u^{2}}=\frac{m}{2}\meanx{\left(\delta v\right)^{2}}+\frac{\hbar^{2}}{8m\sigma^{2}}\:,\label{eq:46}
\end{equation}
and thus at the initial time, where $v=0$:
\begin{equation}
\delta\meanx{E_{\mathrm{kin}}(0)}=0+\left.\frac{m}{2}\meanx{u^{2}}\right|_{t=0}=:\frac{m}{2}u_{0}^{2}=\frac{\hbar^{2}}{8m\sigma_{0}^{2}}\:.\label{eq:47}
\end{equation}
Now we again make use of the equipartition equation Eq.~(\ref{eq:3.9}).
Together with Eq.~(\ref{eq:3.12}) we obtain

\begin{equation}
\frac{m}{2}u_{0}^{2}=\frac{kT_{0}}{2}=\frac{D\zeta m}{2}\:.\label{eq:3.7-1}
\end{equation}
At the time $t=0$ the system is in the prepared state where the fluctuating
kinetic energy term is solely determined by the initial value $\sigma_{0}$,
whereas for later times $t$ it decomposes into the term representing
the particle\textquoteright{}s changed kinetic energy and the term
including $\sigma\left(t\right)$. At $t=0$ the velocity $u_{0}$
is determined by the mean displacement $\sigma_{0}$ and the relaxation
time $\zeta^{-1}$ of the walker, Eq.~\eqref{eq:3.1}. By using Eq.~\eqref{eq:3.7-1}
we can write
\begin{equation}
u_{0}=\zeta\sigma_{0}=\frac{D}{\sigma_{0}}.\label{eq:3.7-2}
\end{equation}
With the Gaussian distribution Eq.~(\ref{eq:42}) being a solution
of the diffusion equation, one has for the particle's drift the familiar
relation
\begin{equation}
\meanx{x^{2}\left(t\right)}=\left.\meanx{x^{2}}\right|_{t=0}+2Dt.\label{eq:51}
\end{equation}
However, comparison with Eq.~\eqref{eq:3.11} indicates that \eqref{eq:51}
is only a solution for the particle in its rest frame. From Eqs.\eqref{eq:46}
and \eqref{eq:47} we see that in order to account for a Gaussian
dispersion as in \eqref{eq:46}, the diffusivity must become time
dependent, too. Thus, one must rewrite Eq.~\eqref{eq:51} as 
\begin{equation}
\meanx{x^{2}}=\left.\meanx{x^{2}}\right|_{t=0}+D_{{\rm t}}t.\label{eq:3.10-1}
\end{equation}
In Refs.~\cite{Groessing.2010emergence,Groessing.2011dice,Grossing.2012quantum,Mesa_Pascasio.2012classical}
we show that $D_{{\rm t}}=u_{0}^{2}t$. We have thus connected the diffusion
processes as described by $D_{{\rm t}}$, which we have found through our
model of the bouncer-walker, to the movement of a particle following
a Gaussian distribution. With this, we have derived the elements of
ballistic diffusion from our classical bouncer-walker model and thus
found a missing link between the different approaches of our previous
work \cite{Groessing.2011explan,Groessing.2010emergence}. We can
now expand our thoughts towards interference and look deeper into
the emergent properties of the quantum world, and beyond. 


\providecommand{\href}[2]{#2}
\begingroup\raggedright\endgroup

\end{document}